# Accurate Reaction-Diffusion Operator Splitting on Tetrahedral Meshes for Parallel Stochastic Molecular Simulations

I. Hepburn,[1,2] W. Chen,[1] and E. De Schutter[1,2,a]

[1] *Computational Neuroscience Unit, Okinawa Institute of Science and Technology Graduate University, Onna, Okinawa 904 0495, Japan*

[2] *Theoretical Neurobiology & Neuroengineering, University of Antwerp, 2610 Antwerp, Belgium*

Spatial stochastic molecular simulations in biology are limited by the intense computation required to track molecules in space either in a discrete time or discrete space framework, meaning that the serial limit has already been reached in sub-cellular models. This calls for parallel simulations that can take advantage of the power of modern supercomputers; however exact methods are known to be inherently serial. We introduce an operator splitting implementation for irregular grids with a novel method to improve accuracy, and demonstrate potential for scalable parallel simulations in an initial MPI version. We foresee that this groundwork will enable larger scale, whole-cell stochastic simulations in the near future.

---

[a] Author to whom correspondence should be addressed. Electronic mail: erik@oist.jp

# I. INTRODUCTION

Recent years have seen the rise of detailed spatial stochastic modeling, by voxel-based methods (e.g. STEPS[1], URDME[2], MesoRD[3], NeuroRD[4]) and particle tracking methods (e.g. Smoldyn[5], MCell[6]). Although the field is still in its infancy, the severe computational demands of tracking molecules individually or in small spatial discretizations whilst accurately simulating their movement and interactions means that the serial limit has already been reached. Serial models can take many weeks or even months to run[7], placing restrictions on the scale of the model and results achievable in practical timeframes. Although the embarrassingly parallel solution of distributing different realizations is a useful form of parallelization, it is only a partial solution, and often a speed-up to the serial simulation is also necessary to make runtimes practical.

Based on Gillespie's original Stochastic Simulation Algorithm[8] (SSA) and extended for diffusion, the Inhomogeneous Stochastic Simulation Algorithm (ISSA) or similar exact methods[9] are inherently serial and very little speedup is achieved by direct parallelization[10]. Therefore, it is also broadly recognized that modifications to the exact algorithms of randomized event times are necessary in order to achieve a scalable solution. Although we will touch on an initial parallel Message Passing Interface (MPI) implementation, this paper will focus mostly on deriving an approximate alternative to the ISSA for irregular grids that is inherently parallel, and includes novel modifications on previous methods to improve accuracy.

Attempts at applying approximations to ISSA methods for spatial reaction-diffusion simulation date back to 2006 with the 'Gillespie Multi-Particle' (GMP) method[11], but are not often specifically tailored towards parallel computing per se. The GMP is a rather simple algorithm based on regular grids that has been shown to be inaccurate under some conditions[12]. However, the idea of applying operator splitting to ISSA methods (that is to separate reaction and diffusion algorithms from each other) for biological systems was introduced in this paper and one could say that the GMP forms the basis of all present operator splitting applications today, including the method introduced in this paper and implemented in STEPS.

The idea was enhanced by Lampoudi et al in 2009 with the 'Multinomial Simulation Algorithm' (MSA)[13], which allows multi-step multinomial diffusion. They make the observation of an error in operator splitting methods of diffusion distance by restricting molecules to nearest neighbor, and go to great efforts to derive an algorithm in 1D that allows molecules to travel within a prescribed distance per diffusion application and distributed binomially. This means that molecules can travel further than nearest neighbor per diffusion application, in effect 'hopping over' regions of space. Although the algorithm includes important concepts such as multinomial distribution and is the first to apply net diffusion transfer, it is unfortunately, in their words, 'considerably more complicated' and unimplemented in 2D, and there is no mention at all of 3D systems. The MSA uses the ISSA for reactions, but only allows one reaction anywhere in the system per diffusion update. Although highly accurate, this is very restrictive in terms of performance gain.

Also in 2009 came the first application of operator splitting to irregular subvolumes [14] which was further developed in 2010 [15]. This approach taken by Ferm et al is to allow 3 different solutions to diffusion: continuous deterministic, tau-leap or SSA-based, adapted locally based on an error estimate and separated by Strang-splitting. Their algorithm is based on excellent theoretical work and the implementation is shown to be fast and accurate under some conditions. However, there appears to be a cost to error estimation and it is shown that systems must be stiff for there to be any performance gain, with the algorithm in fact performing slower than the ISSA implementation known as the Next-Subvolume Method (NSM) for non-stiff, low molecule systems. Since the 'SSA' operation also contained a diffusive term, making the initial algorithm inherently serial, the algorithm was further developed in 2014 [16] for parallelization. This algorithm is based on a two half step Lie-Trotter splitting implementation with an adaptive time-step controlled by local error estimates based on the half-step method, and an initial parallel version applied up to 4 cores shows good scaling promise. Of all approaches to date, this algorithm is the most similar to the method we introduce in this paper, but with some important differences as we will describe.

Tau-leaping is an approximate method initially applied to well-mixed reaction systems that reduces computational cost by calculating a small time step 'tau' over which the event propensities may be approximated as constant and the simulation advanced before updating. Marquez-Lago and Burrage [17] introduced spatial tau-leaping and shortly afterwards Rosinelli et al published an alternative spatial tau-leaping implementation [18], combined with a hybrid system whereby diffusion proceeds deterministically. The Marquez-Lago-Burrage algorithm was modified by Iyengar et al to improve numerical accuracy [19], and Koh and Blackwell later developed another spatial tau-leaping implementation where diffusion is based on net transfer between subvolumes[20]. The tau-leaping approach and operator splitting differ in that operator splitting methods are only approximate in terms of restricting diffusion to predetermined update times and there is no approximation to reaction or diffusion propensities.

The literature suggests many alternatives for a potential approximate method implementation in STEPS and in the next sections, where necessary, we will investigate mathematically and practically their accuracy and performance with a view to justifying inclusion or exclusion in our algorithm.

## II. THEORY

We consider the excellent work in the past decade on ISSA approximation and incorporate some of the ideas in our work, but we take a slightly different approach to many before us by tailoring for the specific goal of scalable parallel computation with the goal of approaching whole-cell simulations: In order to achieve this we set out to satisfy three objectives:

1) Recover reaction and diffusion speeds, expected analytically and simulated by the ISSA. Mathematical theory lays the groundwork for this, but we also place a strong emphasis on validation in practical tests and we will introduce some new models for this purpose as well as using some that we have introduced previously[1].

2) To preserve noise in the system, including spatial variability. This will come at a cost, but in our view is an essential feature of stochastic systems that we don't want to lose.

3) Allow good scaling by maximizing computation per communication and minimizing the frequency of communication across processes in a parallel implementation. We want to go to the upper-limit of communication period that our requirements in 1) and 2) allow in order to achieve the best performance we can in the parallel implementation.

We use some ideas from literature that contribute to our goals and reject others that don't, and introduce a novel method to enhance accuracy in the stochastic operator splitting framework.

## A. Multi-Molecule Diffusion

Whereas ISSA approaches simulate every cross-subvolume diffusion event at the precise time that it is calculated to occur, in operator splitting methods multiple diffusion events are aligned to the same time and at regular intervals (Figure 1(a)). This is obviously an approximation to the exact algorithm, yet is clearly necessary to allow an inherently parallel implementation: reaction and diffusion algorithms may be distributed and computed in parallel, with regular cross-process communication consisting of diffusive transfer. We will refer to the period that we advance the simulation during which no diffusive transfer occurs as '$\tau$', though it is important to emphasize that we are not describing a 'tau-leaping' implementation.

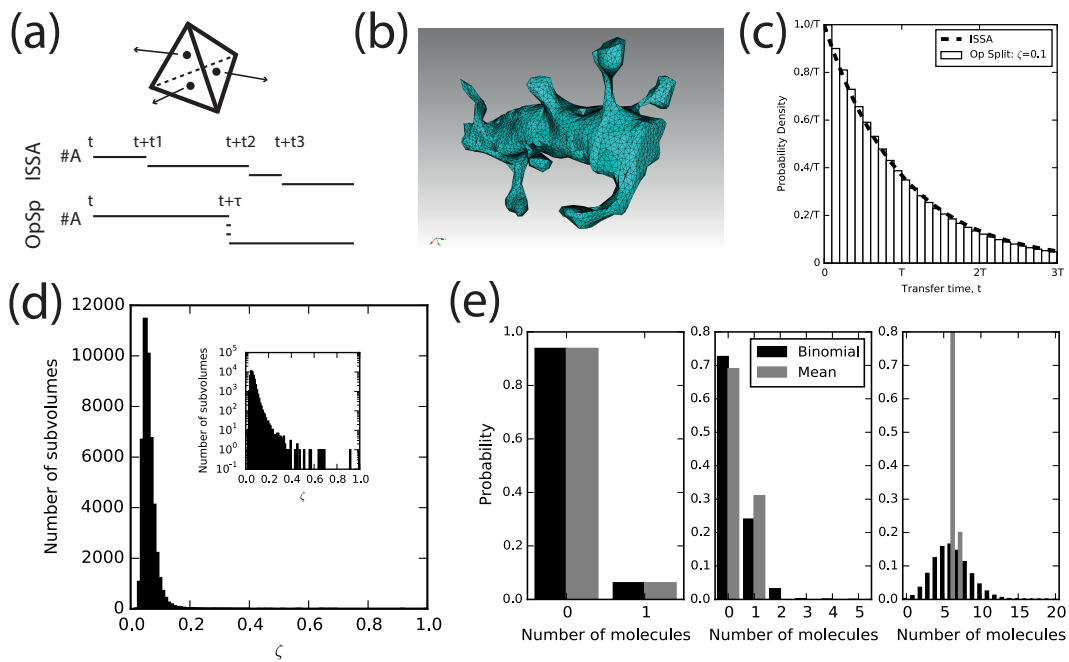

Figure 1: (a) Schematic of the difference between the ISSA and operator splitting for simulating diffusive transfer; in this example the departure of 3 'A' molecules from a tetrahedron to its neighbors. In the ISSA, single molecule events are simulated at unique times, whereas in operator splitting several events are aligned to a predetermined update period, $\tau$. (b) The example spiny dendrite mesh that is analyzed. (c) The probability density distribution of the diffusive transfer period, that is the time each molecule resides within a tetrahedron before diffusing out, for the exact ISSA (black) compared to operator splitting (white bars) for a subvolume where $\zeta=0.1$, that is the local diffusion rate of a molecule in that tetrahedron is 10% of the largest local rate across all tetrahedrons in the mesh (see text). Note: although in operator splitting, transfer takes place at discrete times (right edges of bars), the **displayed bars are of full width for visual comparison. (d) The distribution of ζ for all tetrahedrons in the mesh shown in (b). (e) Comparison of the number of molecules transferred at each step if binomial (black) or the discrete mean (gray) for 3 different example molecule numbers.**

With regards to simulating diffusion in the operator splitting framework, two important concepts must be considered: 1) the number of molecules to diffuse to neighbors per subvolume at **τ**, and 2) how they are distributed to the neighboring subvolumes. For regular grids and with spatially uniform diffusion rates as in the GMP[11], **τ** can be chosen to align precisely to the mean diffusion rate for each type of molecule, and *every* molecule in the system of a certain type then diffuses on its scheduled transfer time. For irregular grids, a slightly different approach is required because of all the varied local diffusion rates due to the different sizes of the tetrahedral subvolumes (Figure 1). Considering the diffusion probability density function (PDF) that underlies each molecule's behavior in exact ISSA methods (Figure 1(c)), one can calculate the probability of a molecule transferring from its subvolume to a neighbor by simply solving the integral in units of **T** = 1/d, where d is the local diffusion rate (in units of s$^{-1}$):

$$p = \int_0^\zeta \left( \frac{1}{T} e^{-\frac{t}{T}} \right) = 1 - e^{-\zeta} \tag{1}$$

where ζ is some fraction of **T**, i.e. ζ = **τ**/**T**. Equation 1 is the cumulative density function (CDF). If ζ=1 and so we are at the mean transfer time for the molecule, then p = 0.632, i.e. the probability that a molecule would move by diffusion in the ISSA within time **T** is equal to 63.2%. So if we apply diffusion at time **T** within the operator splitting framework one might expect that we should transfer 63.2% of the molecules, but that is not the case. Instead we should move more molecules to avoid slowing of diffusion; that is some molecule transfers are being delayed until **T** whereas by the ISSA they would have occurred earlier (Figure 1(a)), and that delay must be compensated for. By this argument one may calculate a 'slowing factor' from the imposed delay at ζ in the exact distribution, which is equivalent to $\zeta - (1 - e^{-\zeta})$ and must be compensated for by sampling an additional time of the exact distribution, **τ₂**, where the extra sample is: $(1 - e^{-\frac{\tau_2}{T}}) - (1 - e^{-\zeta})$. By comparison, this leads to $\zeta = (1 - e^{-\tau_2})$. Then the total probability of a molecule transferring at time ζ**T** in operator splitting is the CDF to **τ₂**: $= (1 - e^{-\tau_2}) = \zeta$. This is just one argument to show that, to avoid slowing, the probability of a molecule transferring after time $\zeta T$ is simply equal to $\zeta$, and since probabilities above 1 are not feasible $\zeta=1$ is the upper limit (above this it is no longer possible to reproduce the correct diffusion speeds). Such transfer calculations have been applied before in operator splitting methods, but this is the rationality for doing so.

For regular grids, as applied for example in the GMP, $\zeta$ can always be chosen to equal 1 for any given species since local diffusion rates are equal everywhere, but for irregular grids $\zeta$ will vary throughout the geometry (Figure 1(d)), and must be constrained to be less than or equal to 1 everywhere. In effect, the upper limit for $\tau$ is that $\zeta$ is equal to 1 for the fastest diffusing species in the smallest tetrahedron, and will be lower than 1 everywhere else. In the example mesh we have presented in previous studies[1] and shown in Figure 1(b), $\zeta$ takes a mean value of 0.06 for the fastest diffusing species (and will be lower for all other chemical species). Happily, this value is comparable to the value of 0.1 found by Lampoudi et al to give an acceptable error by diffusion in their 1D-MSA methods[13]. Indeed 43260 out of 45661 tetrahedrons in the sample mesh- approximately 95%- produce a $\zeta$ value less than 0.1 (Figure 1(d)), so without any special treatment this mesh can intrinsically be assumed to produce accurate diffusion within the operator splitting framework.

There is no guarantee that every tetrahedral mesh will produce acceptable values of $\zeta$. For some meshes $\zeta$ may be too small for most tetrahedrons (too irregular), affecting performance (but not accuracy). For others $\zeta$ may be too large (too regular). The latter case is easily fixed by reducing $\zeta$ everywhere by some factor (reducing the update period); the former case may be dealt with by implementing an approximation that ignores outliers (effectively allowing $\zeta>1$ in those outliers) with a small cost to accuracy when those small tetrahedrons become occupied. The inset in Figure 1(d) demonstrates this idea: as can clearly be seen, the numbers of tetrahedrons where $\zeta$ is close to 1 are very few, and only approximately 10 of 45661 tetrahedrons take a $\zeta$ value greater than 0.4. While, without any special treatment, the mean value of $\zeta$ is acceptable, the best solution may be to ignore 10 or so outliers with highest $\zeta$, effectively allowing the mean $\zeta$ for the rest of the tetrahedrons to increase by approximately a factor of 2.5 and improving simulation performance at a very small cost to accuracy. Such considerations will be different for every different mesh and an analysis as applied in this paragraph would be beneficial in terms of producing optimum performance. It is worth noting at this stage that, although the lower the value of $\zeta$ the better the true PDF is sampled (Figure 1(c) shows $\zeta=0.1$) and so the more accurate the simulation, we will show later however, that regardless of the value of $\zeta$, it is the multinomial distribution of the transfer molecules that is important in faithfully representing spatial noise in the system, not the precise individual timing of transfer events.

With the probability of diffusion equivalent to the local $\zeta$ (which varies with location and each chemical species' diffusion rate) and so this probability can be represented by $\zeta$ for any molecule in any region of the mesh, it is trivial to show that for a population $N$ of some molecule within some subvolume, the number of molecules, $n$, that it is necessary to transfer at diffusion update time $\tau$ is chosen by sampling the binomial distribution: $n$=binomial($N$, $\zeta$). For previous operator splitting implementations it appears sometimes this $n$ is indeed the binomial variable, but other times it is simply the (rounded) mean. The two are only exactly equivalent when $N$=1 but quite different for large $N$ (Figure 1(e)); that is the mean $n$ ($= N\zeta$) is a good assumption for low concentration systems, but for large concentration systems the binomial distribution should be used so as to avoid an error resulting in a possible under-sampling of noise (Figure 2(d): compare 'Gross d, Binomial n' to 'Gross d, Mean n'). Treating molecules individually, and choosing for each whether to transfer or not based on a probability equivalent to $\zeta$, would result in an exact sampling of the binomial distribution, however it is clearly not computationally desirable to take such an approach for large $N$ and instead a standard binomial function can be applied to the population.

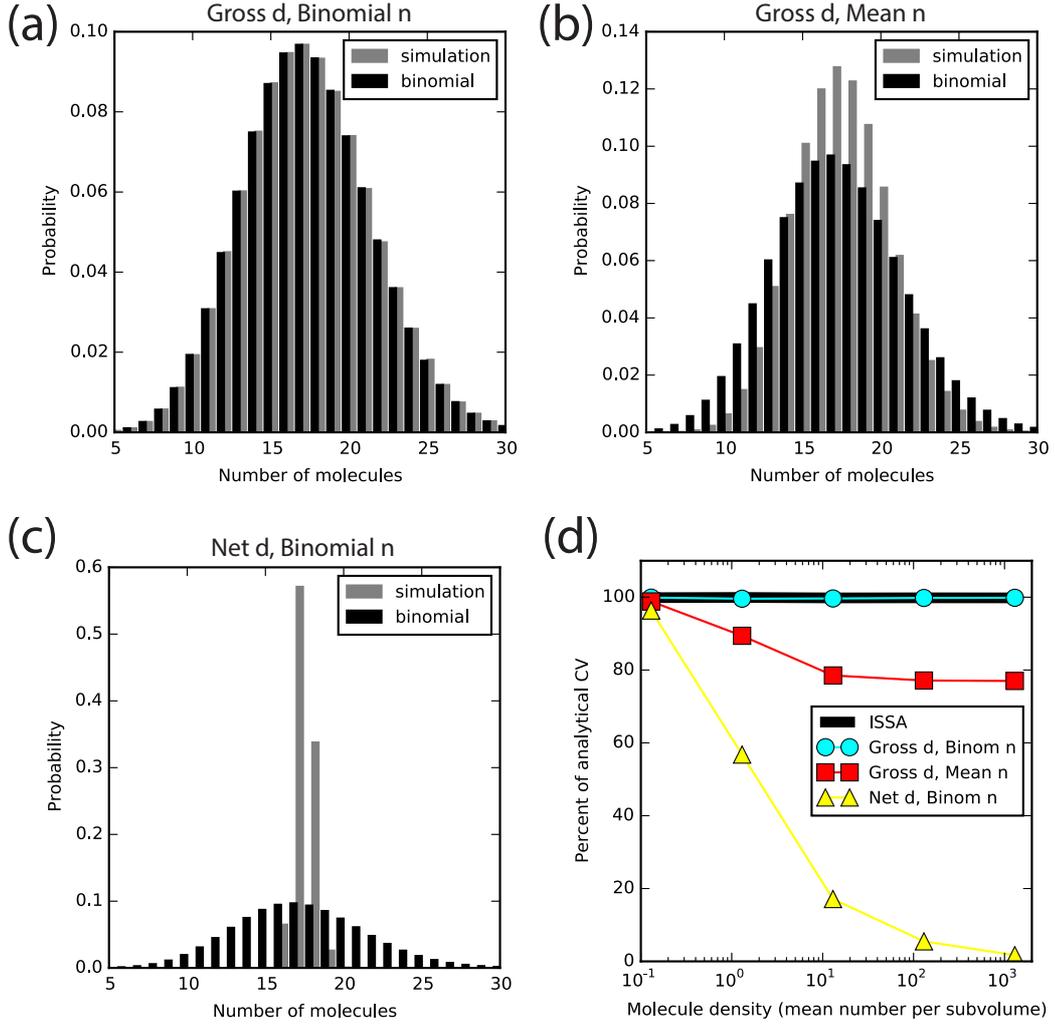

**Figure 2:** Spatial noise simulated by different operator splitting methods. (a),(b),(c) For a diffusion simulation in a tetrahedral mesh with 1*μ*m x 1*μ*m x 1*μ*m cubic boundaries and mean ζ for tetrahedrons = 0.5, for an example tetrahedron the simulated molecule distribution (gray bars) compared to the correct binomial distribution (black bars) if the number of molecules by diffusive transfer, *n*, is (a) binomial and based on gross diffusive flux, (b) the discrete mean calculated by gross diffusive flux, or (c) the number of molecules is binomial but based on net diffusive flux. (d) Comparison to the analytical coefficient of variation of the three simulated methods shown in (a)(b)(c) (shapes) and the ISSA (black line) with varying system size.

A final consideration for choosing *n* is the possibility of simulating only the net diffusive transfer between subvolumes. This approximation has been considered before[13, 20] and the intuitive notion that this reduces spatial noise has been demonstrated before in regular grids but is stated as an accepted loss of accuracy due to the gain in performance [20]. We include a brief analysis of how net diffusion may affect a spatial stochastic simulation in irregular grids. Figure 2(d) demonstrates the drastic loss of spatial noise by net diffusive transfer for all systems expect those of very low molecule density (fewer than 1 molecule per subvolume on average), measured as simulated coefficient of variation of the stationary distribution averaged across subvolumes and compared to the expected average coefficient of variation of the binomial distribution (Figure 2(a),(b),(c) show one example subvolume). With only 1 molecule on average per tetrahedron the spatial variance is reduced to approximately 60% of the ISSA, for 10 molecules per subvolume on average the spatial variance is reduced to just 20% of the ISSA, and reduces even more dramatically for higher densities (Figure 2(d)). These results show that simulating net diffusive transfer is not desirable if one wants to capture spatial noise in the stochastic system.

**B. Molecule Distribution**

Another possible approximation is to distribute molecules uniformly, as opposed to multinomially[13]. We tested a uniform implementation, where each direction was still weighted correctly for the irregular mesh. Uniform molecule number per direction is usually a fractional number, which was randomly rounded up or down so that the correct *n* was applied and in each direction always to the next integer above or below the fractional uniform number. We tested and compared the uniform and multinomial distribution within the same simple model system. As can be seen in Figure 3(a), a uniform distribution does not faithfully capture spatial noise in the system as compared to the analytical solution and the ISSA. Figure 2 and 3 also demonstrate an important finding: spatial noise arises from the randomized, multinomial diffusive transfer between subvolumes and not the precise timing of the transfer as captured by the ISSA. The multinomial algorithm in the operator splitting framework closely reproduces the output of the ISSA and agrees with the analytical solution to within 1%, even when a large *n* is transferred per subvolume per update as in the high molecule density simulations. Therefore, this alignment of diffusion events within the operator splitting framework is a very accurate approximation provided that net transfer is not used, *n* is binomial, and distribution is multinomial.

The results in sections A and B are from a mesh with a mean $\zeta = 0.5$, and there may be some influence of the amplitude of the errors show with different values of $\zeta$. But in general, we assume the results are valid with reasonable ranges of $\zeta$. For some approximations and models, convergence may occur but only with very low $\zeta$ in high molecule density systems at a severe cost to simulation performance. Some solutions may never converge, such as net diffusion with high molecule number.

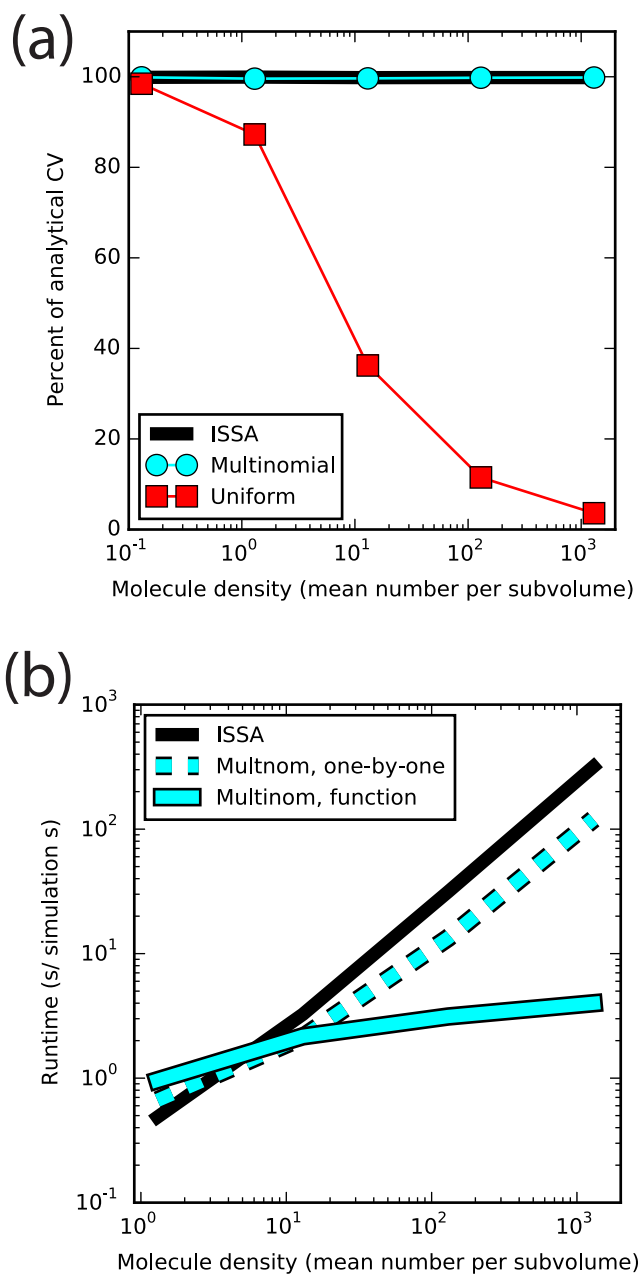

Figure 3: Accuracy and performance of the multinomial operator splitting algorithm. (a) For the same model system as shown in Figure 2, a comparison to the mean expected coefficient of variation across all subvolumes for the ISSA (black line) and operator splitting implementations where molecules are distributed multinomially (circles) or uniformly (squares). (b) Runtimes when the multinomial distribution of molecules is performed individually for each molecule (light dashed line) or by a multinomial function (light solid line) and compared to the ISSA (black line).

Multinomial distribution can be captured either by randomly selecting a direction for each molecule one by one ($n$ random numbers required), or applying a multinomial function that distributes $n$ over the (maximum) 4 neighbors of each tetrahedron (4 or fewer random numbers required). The multinomial function calculates:

the number of molecules out of a total number of molecules to transfer, *n*, from the subvolume calculated for direction *i*,

$$n_i = binomial(n - \sum_{j=0}^{i-1} n_j, p_i),$$

where $p_i$, the probability for direction *i* is:

$$p_i = \frac{w_i}{1 - \sum_{j=0}^{i-1} w_j},$$

where $w_i$ is the 'weight' for the direction in the irregular subvolume and the sum of all weights over the (maximum) 4 directions is equal to 1. $p_i$ always equals 1 for the last direction and so any remaining molecules are transferred in the final direction and the sum of all $n_i$ is equal to n. An alternative is to take *n* as the total population of the tetrahedron, *n=N*, and reduce the weights per direction by the factor ζ discussed in the previous section, to give the same result.

There is a threshold at which the function will perform faster than distributing molecules one by one (Figure 3(b)), which suggests the implementation should incorporate a dynamic switch between the two methods based on local conditions. This threshold is estimated as *n*=10 in our implementation. Figure 3(b) also demonstrates that there is already a performance benefit of the multinomial algorithm in the operator splitting implementation in serial when compared to the ISSA except for very low molecule densities.

## C. Reaction-diffusion with τ-occupancy

Reaction-diffusion operator splitting entails alternating some algorithm for diffusion and another for reactions, and there have been several approaches to this in the past. One approach is to apply a diffusion update after every single reaction chosen by the SSA, effectively reducing τ to the inverse of the zero reaction propensity. The method then removes the reactants before applying the diffusion update, and inputs them upon its conclusion. Whilst one might expect this to be highly accurate, one may also expect a severe performance penalty by reducing τ so drastically, and this does not fit in with our goals. Diffusion events in most models, by design, outnumber reaction events by typically 2 to 3 orders of magnitude yet assuming a mean value for ζ of 0.1 (Figure 1(d)) in the operator splitting framework, then effectively 10% of molecules transfer per diffusion update meaning a great many individual diffusion events occur at each update and, effectively, the period between individual reactions can become smaller than $\tau$ in the operator splitting framework. For example, if 10% of all diffusion events are effectively simulated together at every update and there are 10,000 molecules in the system, then effectively 0.1x10,000 = 1000 individual diffusion events are aligned at every $\tau$. Reaction periods will then be smaller than $\tau$ if reaction events accounted for just 1/1000 = 0.1 percent of all events (both reaction and diffusion) in the ISSA. In a published model[7] we calculate approximately a 250x reduction in $\tau$ if it is based on individual reaction events; in other words, if $\tau$ is based on diffusion rates, one might expect ~250 reactions to occur anywhere in the system between diffusion updates, so clearly it is not desirable to base $\tau$ on the inverse of the zero reaction propensity in terms of performance. Some previous methods do allow for multiple reactions per diffusion update, such as in the GMP[11], though often there don't seem to be any special steps taken to consider how diffusion and reactions affect each other within the operator splitting framework. In our method we want to keep $\tau$ at the upper limit calculated for diffusion, allowing multiple reactions per update, but also to minimize loss of accuracy by considering how the reactions affect diffusion probabilities locally. This differs from previous methods where a half-step method is applied and tau adapted to a tolerance of errors calculated from the half-step application[16].

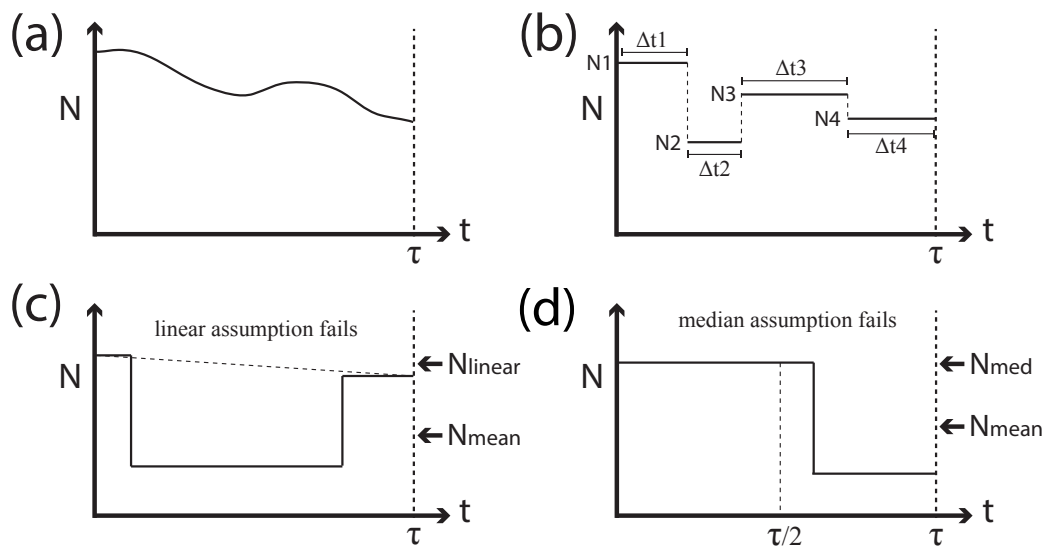

**Figure 4: Schematic of the reaction-diffusion behavior in a subvolume during operator splitting.** The diffusion algorithm is applied at $\tau$ and so $N$, the number of molecules in the subvolume, changes only by reactions until $\tau$. In the continuous case (a) and discrete stochastic case (b), the number of molecules to transfer at $\tau$ depends on the mean

**number of molecules present during τ. (c) and (d) show, respectively, where the linear assumption and the median assumption to approximate the mean *N* during τ fail.**

When allowing multiple reactions to occur between diffusion updates, the problem becomes how to calculate the number of molecules to diffuse to a neighbor at τ, *given that the population may have changed during τ by reactions*. The schematics in Figure 4 demonstrate the problem graphically. Starting with a hypothetical continuous case (Figure 4(a)), some population *N* of a molecule present in a subvolume for small time *dt* with local diffusion rate per molecule of *d* should contribute *N(t).dt.d* to a loss by diffusion during τ. The total loss of molecules during τ due to diffusion, *n*, is therefore:

$$n = \int_0^\tau N(t).dt.d = \overline{N}.\tau.d \qquad (2)$$

The value $\tau.d \leq 1$ and is equivalent to the value defined as ζ in the previous sections (see equation 1).

For stochastic systems (Figure 4(b)), *N(t)* is simply a discretization of the continuous case where the transfer may be written as:

$$n = \sum_0^\tau N_j \Delta t_j d$$

$$= d(N_1 \Delta t_1 + N_2 \Delta t_2 ...)$$

$$= \overline{N}\tau d \text{, as expected.}$$

And so a correction to the transfer calculated in the diffusion-only scenario, as discussed previously, is that the transfer is based on $\overline{N}$, that is the mean *N* during τ. Although a simple result, this is perhaps the most important result we present in this paper, which has not been applied to previous operator splitting methods to the best of our knowledge. When *N* does not change during τ, the transfer is simply *n=N*.ζ, or more correctly, *n*=binomial(*N*,ζ) as derived earlier for diffusion-only systems. When *N* changes during τ the transfer is:

$$n = binomial(\overline{N}, \zeta) \tag{3}$$

This 'mean–corrected binomial distribution' will approximate the variance well in the case that the fluctuations around the mean-occupancy $N$ are small (Sam Yates: personal communication).

Our method, as implemented in STEPS, stores for each subvolume a sequence of $N_j \Delta t_j$ that we call the 'occupancy' and requires, computationally, just two extra floating-point numbers per subvolume: the time since the last change and the cumulative occupancy. Just these two numbers allow STEPS to calculate the occupancy during τ, inexpensively, at every diffusion update.

The algorithm implemented in STEPS is:

---

**Algorithm 1: Descriptive Reaction-Diffusion Operator Splitting algorithm in STEPS**

Precompute τ based on model and geometry properties: τ = min(1.0/$d_{m,tet}$) over all local diffusion coefficients $d$, for all tetrahedrons $tet$, and diffusing species $m$.

while ($t < t_{end}$)

    Step 1: Adjust τ to align to $t_{end}$ if necessary: if ($t+τ > t_{end}$): τ = $t_{end}$-$t$

    Step 2: Run SSA to previous reaction before τ:

        Step 2.1: When each SSA reaction occurs, for all products and reactants update local occupancy.

    Step 3: Run diffusion algorithm at τ:

        For each diffusing species:

            Step 3.1: calculate $\overline{N}$ based on occupancy

            Step 3.2: calculate the number to diffuse to neighbors:

            $n = binomial(\overline{N}, \zeta)$

            Step 3.3: distribute $n$ multinomially

    Step 4: Increase biological time, $t$ += τ

---

Figure 4(c) and 4(d) demonstrate two occasions where, for stochastic simulation, the linear assumption or the median approximation (*N(t)* at τ/2) - effectively what is calculated with half-step splitting methods such as second order Strang splitting - to approximate $\overline{N}$ differ significantly from the true value of $\overline{N}$, so we believe our method is *at least as accurate* as such methods. Any other methods for calculating the transfer that do not approximate $\overline{N}$, such as simply using the *N* at τ (as is employed in some previous operator splitting implementations, e.g. [21]), are less accurate (see Results, Figure 7).

## III. RESULTS

### A. Diffusion Validation

The first results we present are based on our validation set [1], which largely focus on testing the speed of diffusion within different geometrical settings both for volume diffusion (transfer between tetrahedral subvolumes) and surface diffusion (transfer between surface triangles). We have already presented high accuracy with respect to capturing the spatial variance (Figures 2,3). Figure 5 shows example output for each volume diffusion model as we have described previously, with additional models for surface diffusion as supported in STEPS since version 2.0. In stochastic systems any validation has a chance to fail. However, no increased probability of failure was detected for our operator splitting method, which was always less than 1% as for the exact ISSA implementation.

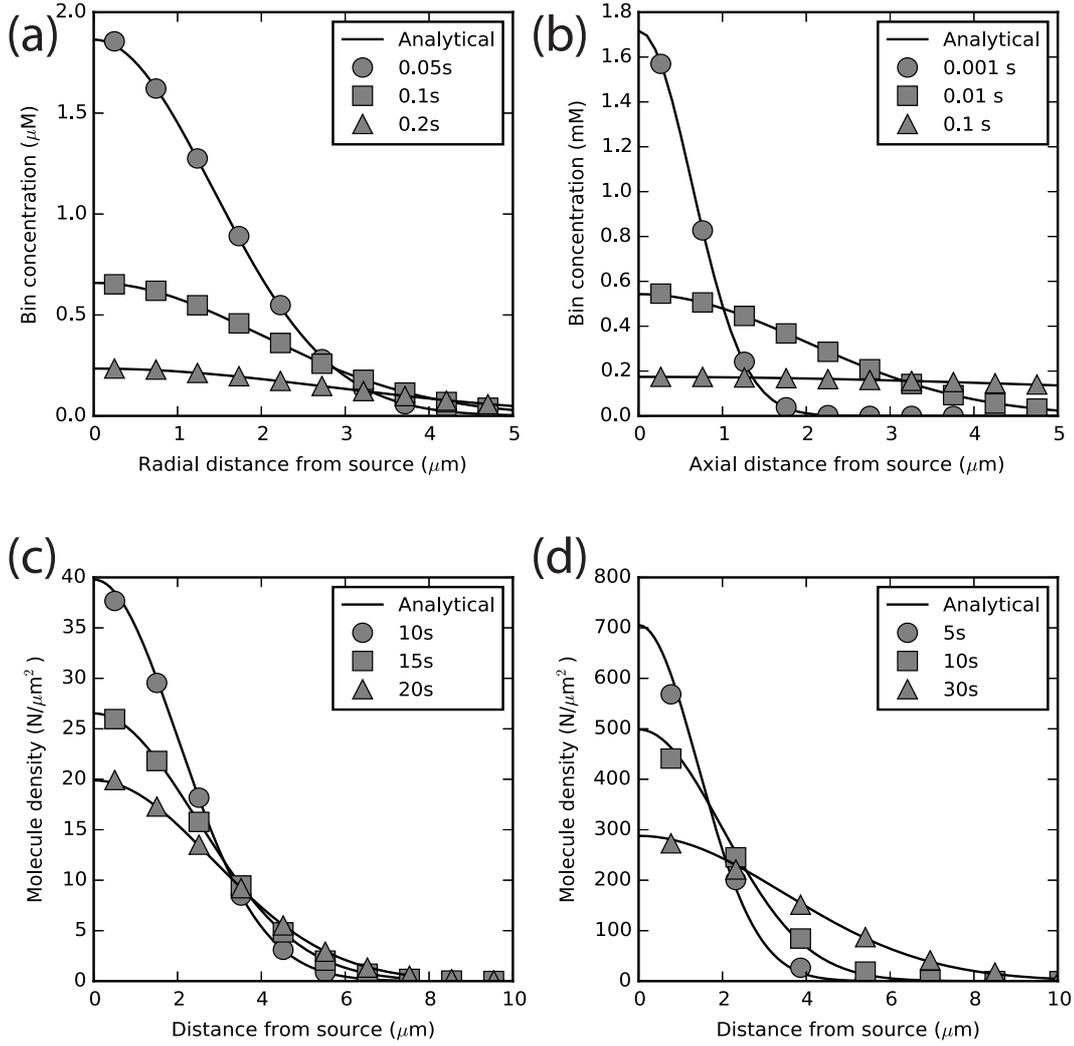

Figure 5: Diffusion speed validation of the operator splitting implementation in STEPS. In each case the simulated model is compared to an analytical solution, with many spatial stochastic runs taken to find the mean behavior, for 3 different simulation times. (a) unbounded volume diffusion, (b) bounded volume diffusion, (c) unbounded surface diffusion and (d) unbounded surface diffusion with a line source. In each case the 'source' of molecules is at distance 0.

## B. Reaction-Diffusion Validation

One of our reaction-diffusion models allows comparison to an analytical solution to the reaction-diffusion master equation [1] and merits deeper investigation, where the expected distribution can be compared to the simulated distribution and the error faithfully measured, both for the ISSA and operator splitting method presented in this paper. The model importantly contains a 2$^{nd}$ order reaction, which is sensitive to the specific reaction-diffusion method and underlying algorithm. We varied system size by varying the production rate constant from a sparse system with less than 1 molecule per tetrahedron on average to a system with multiple molecules on average for every tetrahedron. In each case the analytical solution was fitted to the resulting distribution and an apparent $2^{nd}$ order reaction rate recovered (Figure 6(a),(b) and see [1]). Figure 6 (c) and (d) show how the correct reaction rate was simulated to within an error of 1% for all system sizes, and the error in the operator splitting implementation was comparable to the small error in the ISSA.

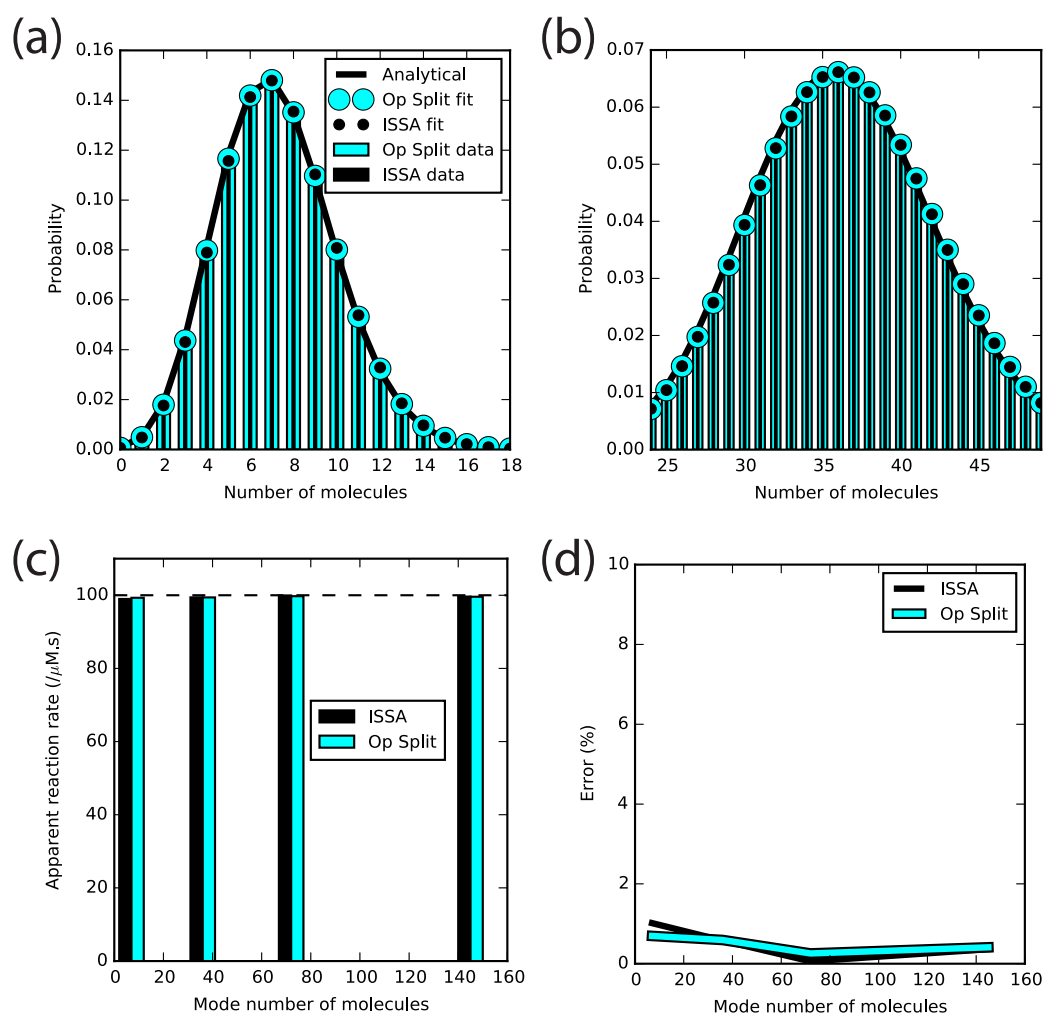

**Figure 6: Reaction-diffusion 'production and $2^{nd}$ order degradation' validation model of the operator splitting implementation in STEPS. (a) and (b)** For two example system sizes (controlled by the production reaction rate) the molecule distribution in the whole system is compared to the analytical solution to the Reaction-Diffusion Master Equation (black line) for the ISSA and operator splitting implementations, both for raw data (bars) and fit to the analytical equation (circles). **(c)** The simulated $2^{nd}$ order reaction rate for ISSA and operator splitting measured from the fits is compared to the expected reaction rate (black dashed line, 100/$\mu$M.s). **(d)** The error in the simulated reaction rate by both methods for different mode (peak) molecule numbers in the system.

Another reaction-diffusion model with a strong spatial element is based on an analytical system '1D diffusion in a finite tube with constant influx at both ends'[1]. The STEPS model is simulated on a cylindrical mesh and contains a stochastic reaction that is restricted to the ends of the cylinder and provides a constant influx of molecules (Figure 7(a)). The molecules diffuse and so, after time, a spatial gradient occurs (Figure 7(b)) for which there is an analytical solution[1]. This model contains a strong interplay between reaction and diffusion with diffusion speed strongly dependent on the production rate and so is a good test of our operator splitting method. Figure 7 compares output from the exact ISSA implementation in STEPS with the operator splitting method introduced in this paper, and also a first order splitting method which differs from our method only in that no special consideration is given to calculate $N$ for the diffusive transfer, and is simply the $N$ at time $\tau$ (Figure 4). Although in this model the error is quite low for all methods, Figure 7(c) and (d) clearly show a larger error for the first order method compared to our operator splitting method. The error in our operator splitting method itself is comparable to the error in the ISSA, the small error in which is presumably largely due to the finite number of iterations that were averaged (5000).

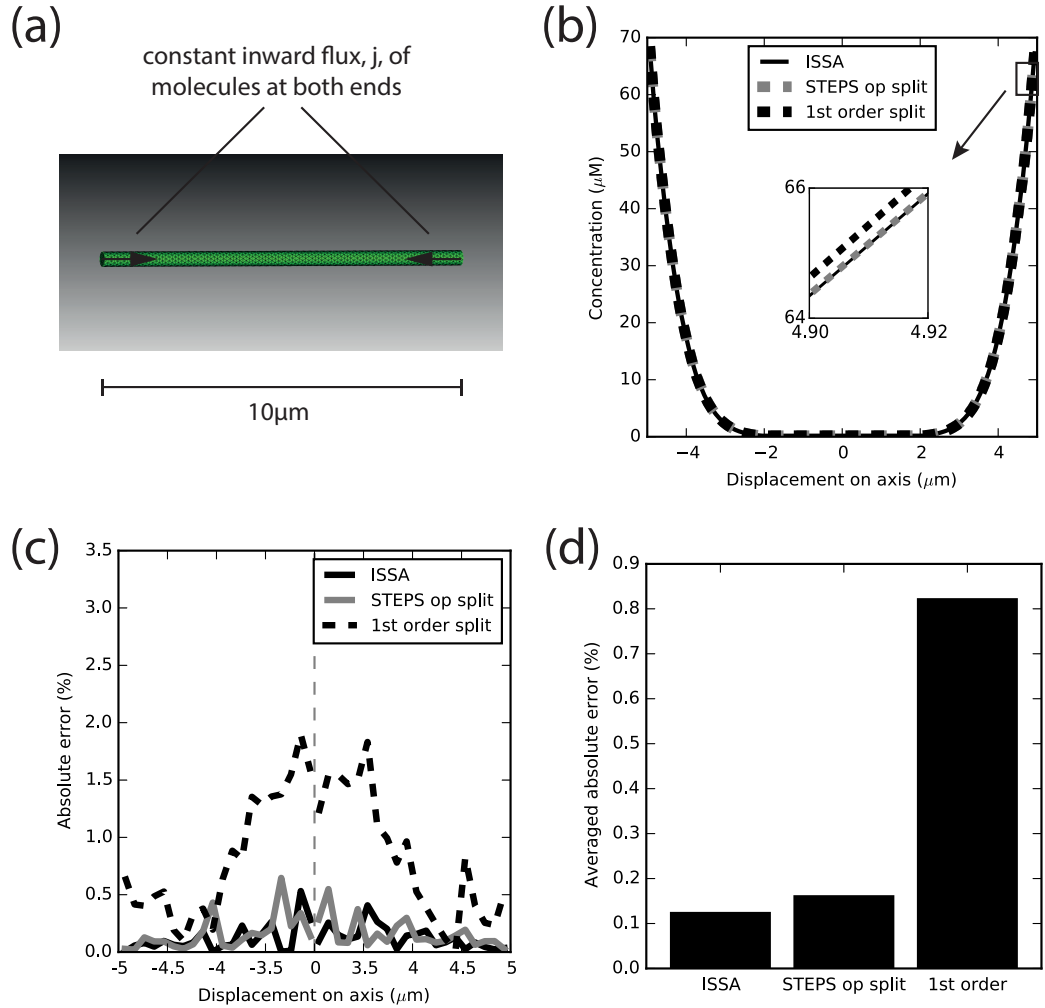

Figure 7: Reaction-diffusion '1D diffusion in a finite tube with constant influx at both ends' validation model of the operator splitting implementation in STEPS. (a) A stochastic production reaction of constant rate is restricted to the tetrahedrons that border the two faces of the cylinder, and the molecules diffuse towards the center. (b) Example concentration after some time along the axis of the cylinder by three methods: the ISSA (black line), operator splitting method in STEPS as described in this paper (gray dashed line) and a first order implementation which differs from the STEPS method only that the $n$ for diffusion transfer at $\tau$ is based on the $N$ at $\tau$ and not the mean $N$ during $\tau$ (see Figure 4 and the description in the text) (black dashed line). (c) The error along the axis at the time shown in (b) for the 3 methods as compared to the deterministic simulation. Note: the absolute error for the first order method appears to pass through zero due to a switch in the sign of the error. (d) The averaged error for all spatial bins of the data shown in (c).

## C. Recovery of Reaction Probability Density Functions

An important consideration for any operator splitting implementation is how well it captures the noise compared to exact stochastic methods. The underlying goal of stochastic simulation is to faithfully capture noise in the biological system and any approach that fails to do this should be rejected. In order to investigate the intuitive belief that our implementation does indeed accurately capture reactive noise, we implemented a reaction-diffusion model modified from a previous study [1] that contains 10 distinct chemical species diffusing between the range of 10 to 100$\mu m^2$/s and reacting by 8 different channels (consisting of 4 reversible reactions) with a broad range of reaction constants from 1-1000/$\mu$Ms for the 4 bimolecular, second-order reactions and 1-100/s for the unimolecular, first-order reactions.

The simulation was run with the operator splitting algorithm in STEPS for 500s with the period between every single reaction event recorded for the 8 different types of reaction. After approximately 10s the system reached a steady state (Figure 8(a)) and so the analytical PDF for each reaction based on the mean molecule number could be calculated and compared to the recorded probability density. In every case the simulation captured the expected probability density very closely- Figure 8(b) shows one example: Reaction #1 A+B > C bimolecular reaction. The simulated probability density could be fit to an exponential distribution with a free rate parameter, and the rate from the best fit compared to the expected rate. Figure 8(c) shows an example least squares fit, again for Reaction #1, and Figure 8(d) shows the comparison of the fits to the expected rates for all 8 reactions. In each case the simulated rate was recovered to within a 0.5% error and the one standard deviation errors are small indicating a good fit to the data and therefore full recovery of the expected probability density. To our knowledge, no other operator splitting implementation has been proven to capture reaction noise as accurately before.

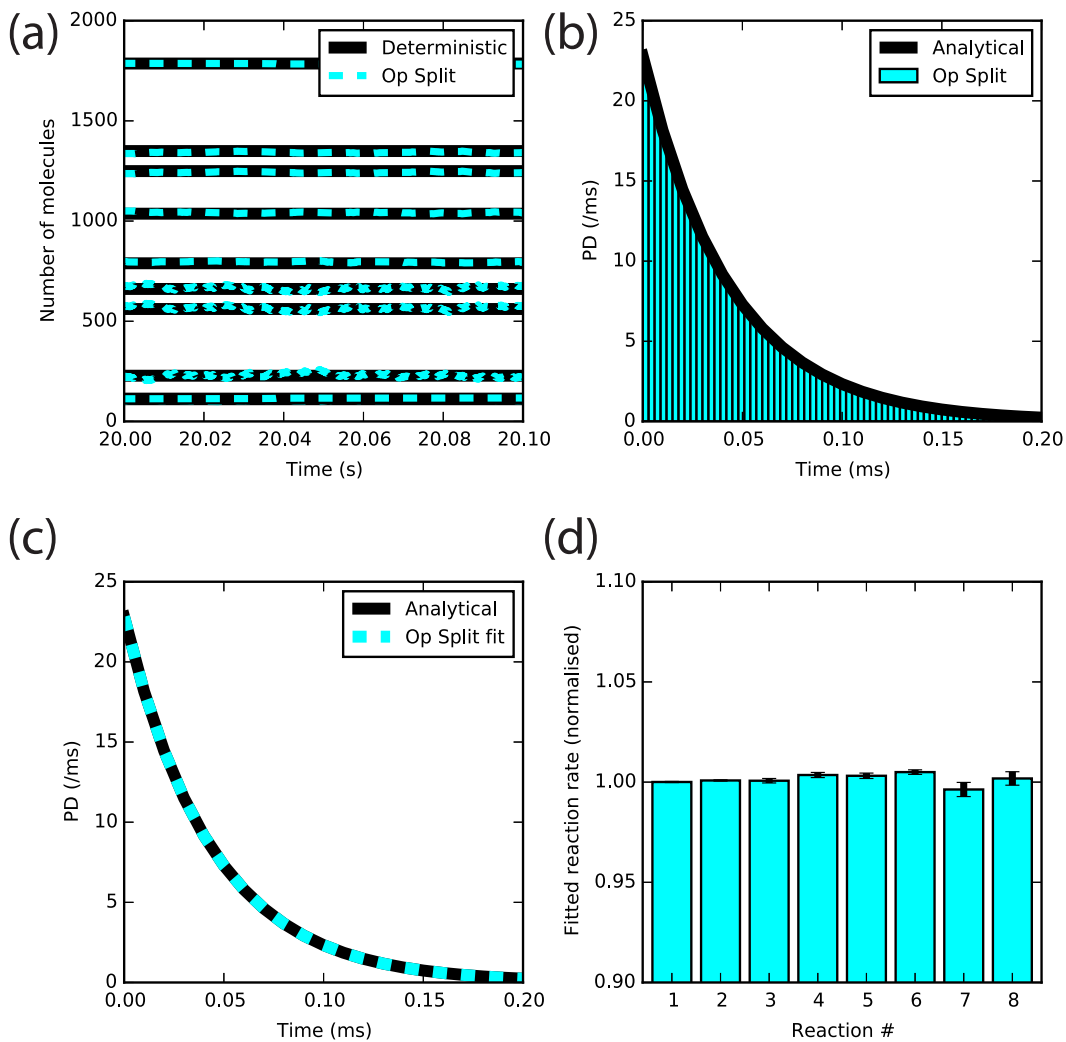

**Figure 8:** Operator Splitting simulation of reaction-diffusion model consisting of 10 chemical species and 8 reactions as described further in the text. (a) Small time window of the full 400s simulation showing that the 10 molecular counts have reached steady state after approximately 10s. (b) The probability density of simulated reaction periods for the A+B > C reaction, in bins of width 0.003ms (bars) compared to the expected exponential distribution (black line). (c) For the same data as shown in (b), a least squares fit to the exponential distribution (light dashed line) compared to the analytical distribution (solid black line). (d) For all 8 reactions, the normalized fitted reaction rate (bars) with one standard deviation of the fit (black error bars) (note: displayed range is 90%-110%).

## D. Initial Parallel Implementation

The operator splitting solution presented in this paper is, by design, friendly to massive parallelization on platforms such as a High Performance Computer (HPC) or a multi-core desktop computer. We created a parallel implementation using the Message Passing Interface (MPI) protocol and used a Mac Pro station with 2.7 GHz 12-Core Intel Xeon E5 CPU and 64GB memory to perform a pilot test of the parallel solution. In this implementation, each MPI process simulates a fraction of the complete geometry. At the initial stage, τ (the period between diffusion applications) is decided globally, as described in Algorithm 1. At the start of an MPI cycle, each process firstly performs reactions by the SSA independently till τ is reached. When τ is reached, the diffusion algorithm described in Algorithm 1 step 3 is applied in each process, whilst remote molecule changes and propensity updates are stored as local buffers. These buffers are then synchronized and applied across affected neighboring processes. This completes the update cycle.

We used three test cases to compare the performance of the parallel operator splitting implementation with the performance of Tetexact, the serial ISSA implementation in STEPS[1]. A cylinder was used for all simulations, with 2μm diameter and 10μm length along the z-axis, consisting of 1060 tetrahedrons, with one face defined at $z=0\mu m$ and the other at $z=10\mu m$. In the parallel simulations, we partitioned and distributed the cylinder along the z-axis, based on the number of processes used. Giving $T_{SSA}$ as the serial ISSA wall-clock runtime for a test case, and $T_{MPI}$ as the parallel operator splitting wall-clock runtime for the same test case with a number $P$ MPI processes, we define speedup for a simulation with $P$ processes as:

$Speedup_N = T_{SSA} / T_{MPI}$

Results of the simulations with $P = 1$ indicate how well serial operator splitting with extra MPI instruments performs against the serial ISSA. We then incrementally increased $P$ from 2 to 10 to study how parallelization affects the performance. Each simulation was repeated 20 times and runtime averaged as the mean wall-clock runtime of the simulation for every $P$.

We first investigated how the parallel implementation performs with a simple diffusion model as a near-ideal circumstance. We uniformly distributed $10^5$ molecules in the cylinder and allowed them to diffuse freely within the cylinder for a given simulation time $\Delta t$. Comparing to the serial ISSA, the operator splitting solution achieves a speedup of 10 times with $P = 1$, thanks to the efficient multinomial approach applied to the diffusion approximation. Further speedup can be achieved by increasing the number of processes so that more diffusion events are executed simultaneously; however there is a decrease of efficiency with higher $P$ because using more processes inevitably increases the amount of cross-process diffusion events. In this test case the performance improved markedly with increase of *P* up to *P*=10 with a speedup of over 45 times compared to the serial ISSA (Figure 9(b)).

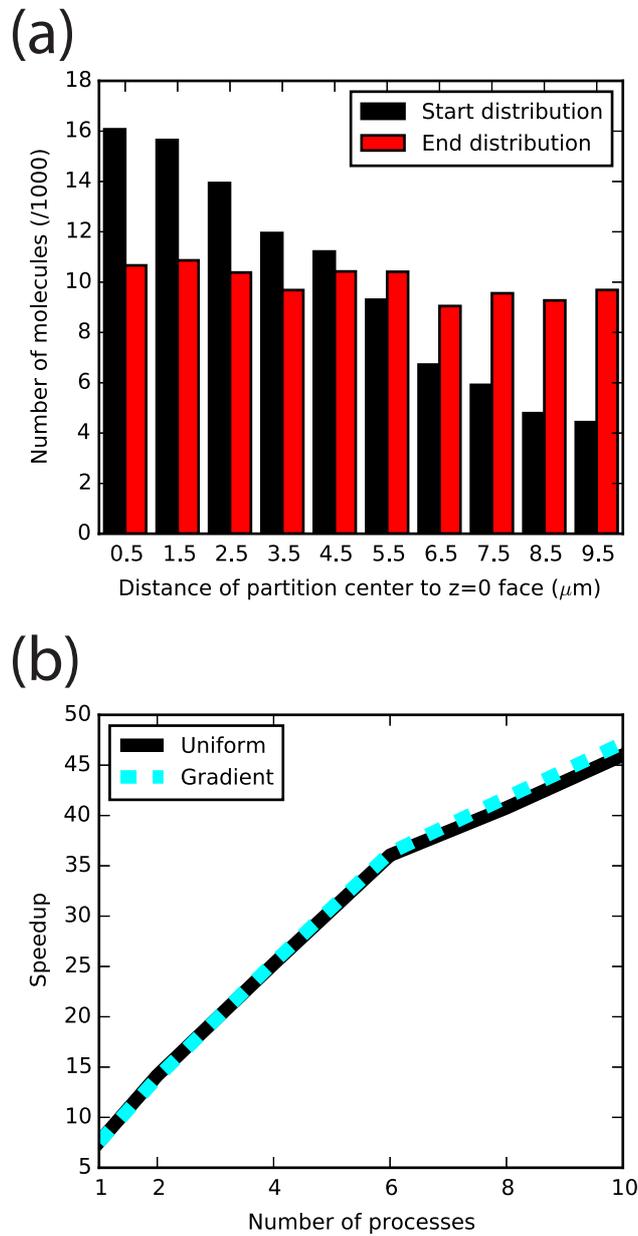

**Figure 9: Diffusion MPI simulation performance. (a) Molecule distributions before and after the gradient simulations in bins of 1$\mu$m width. (b) Speedup of both uniform and gradient diffusion simulations with parallel operator splitting implementation against serial ISSA implementation.**

In realistic research models, molecules are often distributed with a spatial gradient of concentration. In the second test of our MPI implementation, instead of distributing molecules uniformly, we positioned $10^5$ molecules at one end of the cylinder and allowed them to diffuse for a certain period of time so that molecules were distributed with high concentration towards the source end and low concentration towards the other. We continued to run the simulation for $\Delta t$ and recorded the wall-clock time $T_{\Delta t}$ for comparison. Figure 9(a) shows the distribution difference before and after $\Delta t$. As shown in Figure 9(b), a spatial gradient of molecules causes imbalance loading in each process, however the simulations performed almost identically to the uniform model, showing no performance cost despite the uneven load. The reason is that, at this concentration, the multinomial function is almost always applied (eq 2,3 and Figure 3(b)) meaning approximately the same computation is applied for both models even though very different numbers of molecules are being transferred at different regions with a concentration gradient.

We next implemented a reaction-diffusion model as described in the previous section consisting of 10 molecular species and 8 different reactions, in the same cylindrical geometry as for the diffusion-only MPI test. As shown in Figure 10, the parallel implementation achieves similar performance to the serial ISSA with $P=1$, and gains substantial speedup as the number of processes increases, performing approximately 6.5 times faster than the ISSA for $P=10$. Figure 10(b) indicates the scalability of each of the 3 models described, where this time the speedup is compared to the performance at $P=1$ (not the ISSA).

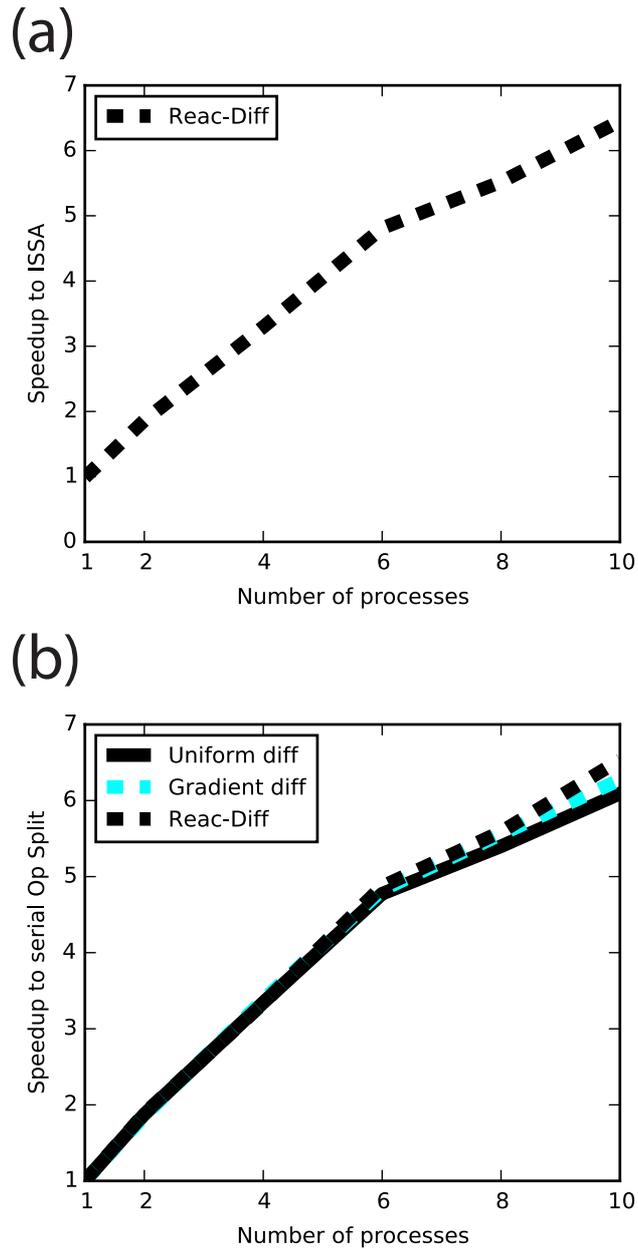

**Figure 10: Reaction-diffusion MPI simulation performance. (a) Speedup of the 10 species reaction-diffusion model described in the text, with parallel operator splitting implementation against serial ISSA implementation. (b) Speedup of all 3 parallel models to the 1 process, serial operator splitting implementation.**

## IV. DISCUSSION

### A. ISSA approximation methods

Based on literature and our own investigations presented in this paper, we briefly discuss the possibilities for ISSA approximation in tetrahedral meshes with particular regard to their suitability of parallel implementation, their strengths and weaknesses, and our justification for inclusion or rejection in our solution for the STEPS software:

### 1. Tau-leaping

Clearly the first step to achieving a scalable parallel solution is to replace the randomized event times in the ISSA (or similar method) with predetermined execution times the simulation can advance to without the necessity of communication by diffusion, which could be cross-process in a parallel implementation. The question becomes rather how to select a tau and which algorithm to implement during tau. The term 'tau-leaping' has become synonymous with a method of advancing a simulation time by an amount 'tau' where propensities are not allowed to change and event frequency calculated on starting propensities. Clearly this fits on the umbrella of predetermined execution times, but comes at the expense of the extra computation required to calculate tau and broadcast at every step in the parallel process. There is also, by design, an error introduced, which is controlled but for low-molecule systems means that tau must be very small, basically reducing the algorithm to the serial ISSA for low molecule systems with worse performance because of the extra cost of calculating tau[20]. In addition, if tau is small and very few events are executed per tau due to many 'critical reactions' then the solution is not scalable. So for reasons of accuracy and performance we decided against a tau-leaping approach.

### 2. Operator Splitting

A scalable solution requires maximizing the computation that takes place on each process between cross-process updates. For arguments that we have presented, the best way to achieve this is to separate reaction from diffusion and deal with them separately, similarly to previous approaches[11, 16, 21].

### 3. Net diffusive Transfer

Although expected to significantly reduce communication, we reject a net diffusion approach because of the drastic cost to spatial noise ([20] and Figure 2(c),(d)), the capture of which is one of the important features of a spatial stochastic simulation. And one should never discount the fact that we are trying to recapture the true behavior of real biological systems, and biology presumably does not deal with net transfer.

### *4. Multinomial distribution*

Multinomial distribution of molecules is essential to maintain spatial noise in comparison to the exact ISSA (Figure 3(a)). For this reason we choose to implement multinomial diffusion. Our diffusion is to nearest neighbor, in part to avoid 'hopping over' regions of space that may be important. Our preference is to track a molecule's exact path through the discretized space, even if at altered leap times in comparison to the exact ISSA, which is unavoidable to achieve a scalable parallel solution. We have shown that our algorithm captures both spatial noise caused by stochastic diffusion (Figures 2, 3) and diffusion speeds (Figures 5, 7) accurately.

### *5. Local error estimates*

Local error estimation only serves to reduce $\tau$ (since we are already at the upper limit) and, in our implementation, that is neither necessary for accuracy nor desirable for performance. $\tau$ is intrinsically kept small anyway in irregular grids (Figure 1), and accuracy can be enhanced by considering how stochastic reactions affect local diffusion propensities (Figure 4, 7). We have proven our method to be accurate under a wide range of conditions without the need for a costly error estimation step, which is important for performance because error estimation can actually make the algorithm perform much slower than the direct ISSA in non-stiff systems [15] (in this cited example almost 20 times slower). In our implementation there is only a small cost to the operator splitting method in serial, which shows as a factor of 2 slowdown compared to the exact ISSA in very low density systems, outperforming the ISSA otherwise (Figure 3(b)), and when parallelized significantly outperforms the ISSA (Figure 9,10).

Although an adaptive $\tau$ could be calculated in parallel, it would also negatively impact performance by the necessity of broadcasting $\tau$ at every step on a parallel implementation (see next).

### *6. Adaptive τ*

Tau-leaping and other approaches allow for adaptive τ. In a parallel setting, adaptive τ will only improve performance if the cost of calculating the τ and broadcasting it is less than gain in performance from potentially calculating a larger τ on average. In our implementation, the τ depends only on the properties of the mesh and the model, and is therefore fixed, reducing the cost of calculating the τ adaptively. Note, however, that in STEPS reaction and diffusion rates may be altered during the simulation, which can affect τ. In such a case τ is recalculated as necessary.

## B. Our operator splitting method

### *1. Accuracy*

We have developed an operator splitting method based on the practicalities of biological reaction-diffusion systems and considering stochastic features that are not present in deterministic systems. Although the field of operator splitting for approximating differential equations is well advanced, it is generally applied with continuous, deterministic solutions in mind. We feel our simple algorithm is well suited to stochastic reaction-diffusion systems. Instead of employing half time step methods that could employ a median or linear assumption that are not necessarily the best assumptions on some occasions for discrete, stochastic systems, (Figure 4) we tailor our method to capture discrete events accurately which can affect diffusion speed (Figure 7). In addition we have shown that it is vital that our method distributes diffusion transfers multinomially and avoids net diffusion to capture spatial noise faithfully (Figure 2, 3). Capturing the correct, stochastic molecule distribution is by far the most important concept for simulating spatial reaction rates correctly and our method is proven to recover reaction rates (Figure 6) and PDFs (Figure 8) accurately under a wide range of realistic biological conditions. We expect that our method is more accurate than splitting methods that to do not consider how stochastic reactions can alter diffusion rates during τ (Figure 4,7).

### *2. Parallel performance*

Overall, the parallel operator splitting implementation provides substantial speedup compared to the serial ISSA implementation, in both diffusion and reaction-diffusion simulations, with both uniform and gradient molecule distributions. For some models of low concentration and with spatial gradients dynamic load balancing may be beneficial, which will however come at a cost in the load balancing calculation. There are many more questions regarding the scalability of the implementation to explore. For example, how well the implementation adapts to a finer mesh, or to a more complex model, or with massive number of processes in a HPC. We plan to address these questions systematically in a future study.

## V. CONCLUSIONS

In this study we have presented an operator splitting implementation for simulating reaction-diffusion systems on irregular grids, with a novel method to enhance accuracy. The method has proven to be accurate with respect to analytical, deterministic and exact stochastic simulations under a wide range of tested conditions. An initial MPI implementation shows good scalability and therefore good promise for speeding up large-scale models in a parallel setting. Future work will include extending this MPI study with an in-depth look at much larger scale systems and their performance when distributed on supercomputers, with expected orders of magnitude speedup in many cases. The theoretical work presented in this paper is therefore expected to contribute to the realization of much larger scale stochastic biological molecule simulations than have been possible to date, significantly enhancing the tools that a researcher has at their disposal to advance our understand of molecular biology and how cells behave in their natural setting, with more detailed computational simulation of their behavior than has ever been achieved before.

**ACKNOWLEDGEMENTS**

This work was funded by the Okinawa Institute of Science and Technology Graduate University.

We are very grateful to Sam Yates of the Blue Brain Project, EPFL, Geneva, for his critical review of the manuscript, which led to several improvements.

**SOFTWARE AVAILABILITY**

Source code for the STEPS simulator and model codes can be obtained from http://steps.sourceforge.net